\documentclass[12pt,thmsa]{article}
\usepackage{amsmath}
\usepackage{graphicx}

\begin{document}

\textbf{AXIOMATIC GEOMETRIC FORMULATION OF}

\textbf{ELECTROMAGNETISM WITH ONLY ONE AXIOM: THE }

\textbf{FIELD EQUATION FOR} \textbf{THE} \textbf{BIVECTOR FIELD\ }$F$\textbf{%
\ WITH}

\textbf{AN EXPLANATION OF THE TROUTON-NOBLE }

\textbf{EXPERIMENT}\bigskip \bigskip

\qquad Tomislav Ivezi\'{c}

\qquad\textit{Ru%
\mbox
{\it{d}\hspace{-.15em}\rule[1.25ex]{.2em}{.04ex}\hspace{-.05em}}er Bo\v
{s}kovi\'{c} Institute, P.O.B. 180, 10002 Zagreb, Croatia}

\textit{\qquad ivezic@irb.hr\bigskip \bigskip }

\noindent In this paper we present an axiomatic, geometric, formulation of
electromagnetism with only one axiom: the field equation for the Faraday
bivector field $F$. This formulation with $F$ field is a self-contained,
complete and consistent formulation that dispenses with either electric and
magnetic fields or the electromagnetic potentials. \emph{All physical
quantities are defined without reference frames,} the absolute quantities,
i.e., they are geometric four dimensional (4D) quantities or, when some
basis is introduced, every quantity is represented as a 4D coordinate-based
geometric quantity comprising both components and a basis. The \emph{new}
observer independent, expressions for the stress-energy vector $T(n)$
(1-vector), the energy density $U$ (scalar), the Poynting vector $S$ and the
momentum density $g$ (1-vectors), the angular momentum density $M$
(bivector) and the Lorentz force $K$ (1-vector) are directly derived from
the field equation for $F$. The local conservation laws are also directly
derived from that field equation. The 1-vector Lagrangian with the $F$ field
as a 4D absolute quantity is presented; \emph{the interaction term is
written in terms of} $F$ \emph{and not, as usual, in terms of} $A$. It is
shown that this geometric formulation is in a full agreement with the
Trouton-Noble experiment. \bigskip \bigskip

\bigskip \bigskip

\noindent \textbf{1. INTRODUCTION\bigskip }

\noindent In the usual Clifford (geometric) algebra treatments of the
classical electromagnetism, e.g., with multivectors [1-3] (for a more
mathematical treatment of the Clifford algebra see also [4]), one starts
with a single field equation using the Faraday bivector field $F$ and the
gradient operator $\partial $ (1-vector), see Eq. (\ref{MEF}) below. In
order to get the more familiar form the bivector field $F$ is expressed (in
[1,2]) in terms of the sum of a relative vector $\mathbf{E}_{H}$
(corresponds to the three-dimensional (3D) electric field vector $\mathbf{E}$%
) and a relative bivector $\gamma _{5}\mathbf{B}_{H}$ ($\mathbf{B}_{H}$
corresponds to the 3D magnetic field vector $\mathbf{B}$, and $\gamma _{5}$
is the (grade-4) pseudoscalar for the standard basis $\left\{ \gamma _{\mu
}\right\} $) by making a space-time split in the $\gamma _{0}$ - frame,
which depends on \emph{the observer velocity }$c\gamma _{0}$; the subscript $%
H$ is for ``Hestenes.'' Both $\mathbf{E}_{H}$ and $\mathbf{B}_{H}$ are, in
fact, bivectors. Then the following relations (from [1-2])
\begin{equation}
F=\mathbf{E}_{H}+c\gamma _{5}\mathbf{B}_{H}\mathbf{,\quad E}_{H}=(F\cdot
\gamma _{0})\gamma _{0},\ \gamma _{5}\mathbf{B}_{H}=(1/c)(F\wedge \gamma
_{0})\gamma _{0}  \label{hf}
\end{equation}
are understood as that they \emph{define} $F$ in terms of the sum of $%
\mathbf{E}_{H}$ and $\gamma _{5}\mathbf{B}_{H}$ and \emph{the components of}
$F$ \emph{are considered to be determined by} $\mathbf{E}_{H}$ \emph{and} $%
\mathbf{B}_{H}$, \emph{i.e., by the components of the 3D }$\mathbf{E}$ \emph{%
and} $\mathbf{B}$. Similarly in [3] $F$ is decomposed in terms of 1-vector $%
\mathbf{E}_{J}$ and a bivector $\mathbf{B}_{J};$ the subscript $J$ is for
``Jancewicz,''
\begin{equation}
F=\gamma _{0}\wedge \mathbf{E}_{J}-c\mathbf{B}_{J},\quad \mathbf{E}%
_{J}=F\cdot \gamma _{0},\ \mathbf{B}_{J}=-(1/c)(F\wedge \gamma _{0})\gamma
_{0}.  \label{J1}
\end{equation}
It is supposed in [1-3] that the right hand sides ($\mathbf{E}_{H}$, $%
\mathbf{B}_{H}$ and $\mathbf{E}_{J}$, $\mathbf{B}_{J}$) of the first
equations in (\ref{hf}) and (\ref{J1}) determine the left hand sides ($F$).
We remark that it is generally accepted in the geometric algebra formalism
(and in the tensor formalism as well) that the usual Maxwell equations (ME)
with the 3D vectors $\mathbf{E}$ and $\mathbf{B}$ and the field equation
written in terms of $F$, Eq. (\ref{MEF}) below, are completely equivalent.
Further both in the tensor formalism, e.g., [5], and in the geometric
algebra formalism it is assumed that the components of the 3D $\mathbf{E}$
and $\mathbf{B}$ define in a unique way the components of $F$ according to
the relations
\begin{equation}
E_{i}=F^{i0},\quad B_{i}=(-1/2c)\varepsilon _{ikl}F_{kl}.  \label{sko1}
\end{equation}
In (\ref{sko1}) the components of the 3D fields $\mathbf{E}$ and $\mathbf{B}$
are written with lowered (generic) subscripts, since they are not the
spatial components of the 4D quantities. This refers to the third-rank
antisymmetric $\varepsilon $ tensor too. The super- and subscripts are used
only on the components of the 4D quantities. Greek indices run from 0 to 3,
while latin indices $i,j,k,l,...$ run from 1 to 3, and they both designate
the components of some geometric object in some system of coordinates. (It
is worth noting that Einstein's fundamental work [6] is the earliest
reference on covariant electrodynamics and on the identification of
components of $F^{\alpha \beta }$ with the components of the 3D $\mathbf{E}$
and $\mathbf{B.}$) All this means that the 3D $\mathbf{E}$ and $\mathbf{B}$
and not the $F$ field are considered as primary quantities for the whole
electromagnetism. Even in the very recent geometric approaches to classical
electrodynamics [7,8] the 3D $\mathbf{E}$ and $\mathbf{B}$ are considered as
primary quantities. Thus it is stated in [7]: ``The electromagnetic field
strength $F_{ij}=(\mathbf{E},\mathbf{B})$ (in [7] $i,j,k,...=0,1,2,3$, my
remark) is composed of the electric and magnetic 3-vector fields.'' In order
to get the wave theory of electromagnetism the vector potential $A$
(1-vector) is usually introduced and $F$ is defined in terms of $A$ as $%
F=\partial \wedge A$. In that case the $F$ field appears as the derived
quantity from the potentials. Thence in almost all usual treatments of the
electromagnetism, both in the tensor formalism and in the geometric algebra
formalism, the theory is presented as that the $F$ field does not have an
independent existence but is defined either by the components of the 3D $%
\mathbf{E}$ and $\mathbf{B}$ or by the components of the electromagnetic
potential $A$. (An exception is, e.g., [9], in which $F$ is an independent
quantity and the 4D $E$ and $B$ are considered as observer dependent
functions of $F$.)

In this paper we present an axiomatic formulation of the classical
electromagnetism that uses the bivector field $F$ (\emph{an observer
independent 4D quantity}) in which only the field equation with $F$ (Eq. (%
\ref{MEF}) below) is postulated. The presented formulation with the $F$
field is a self-contained, complete and consistent formulation that does not
make use either electric and magnetic fields or the electromagnetic
potential $A$ (thus dispensing with the need for the gauge conditions). In
such formulation \emph{the} $F$ \emph{field is the primary quantity for the
whole classical electromagnetism both in the theory and in experiments; }$F$
is a well-defined 4D \emph{measurable} quantity.

In this geometric approach to electromagnetism physical quantities in the 4D
spacetime are represented by Clifford multivectors. They are defined without
reference frames (when no basis has been introduced), 4D absolute quantities
(AQs), or, equivalently, they are written as 4D coordinate-based geometric
quantities (CBGQs) comprising both components and a basis (when some basis
has been introduced). Thus these 4D quantities are independent of the chosen
inertial frame of reference and of the chosen system of coordinates in it,
i.e., they are \emph{observer independent quantities}.

In the field view of particle-to-particle interaction the electrodynamic
interaction between charges is described as two-steps process; first fields
are seen as being generated from their particle sources and then the fields
so generated are perceived as interacting with some target particle. The
description of the first step in the $F$ formulation of electrodynamics is
given in Sec. 2.2. In Sec. 2.3 the general solution for $F$ is applied to
the determination of the electromagnetic field $F$ of a point charge. In
Sec. 2.4 the integral form of the field equation for $F$ is constructed
which is equivalent to the local field equation (\ref{MEF}). The second step
of the description of the interaction process requires the determination of
the Lorentz force in terms of $F$ and its use in Newton's second law. This
is at the same time the way in which the components of $F$ are measured in
the chosen reference frame. It is described in Sec. 2.5. We also give the
\emph{new} expressions for \emph{the observer independent} stress-energy
vector $T(n)$ (1-vector)$,$ the energy density $U$ (scalar, i.e., grade-0
multivector), the Poynting vector $S$ (1-vector)$,$ the angular momentum
density $M$ (bivector) and the Lorentz force $K$ (1-vector). They are all
\emph{directly derived from the postulated field equation with} $F$ (Eq. (%
\ref{MEF})) and presented in Sec. 2.6. The local charge-current density and
local energy-momentum conservation laws are also \emph{directly derived from
that field equation with }$F$ and there is no need to introduce the
Lagrangian and the Noether theorem. These laws are presented in Sec. 2.7.
(Of course the integral conservation laws can be similarly derived but it
will not be done here.) In contrast to our theory with one postulated
equation (Eq. (\ref{MEF})) the recent theory [8] (also a geometric approach)
deals with three postulated equations. It will be shown here that all three
axioms from [8] simply follow from our axiom (\ref{MEF}). Furthermore, it is
considered in [7,8], as in almost all other treatments, that $F$ is
determined by the 3D $\mathbf{E}$ and $\mathbf{B}$. As it is said the
exposed formulation with the field equation for $F$ does not need the
Lagrangian. However in Sec. 3 a brief exposition of the Lagrangian
formulation with the $F$ field is presented; \emph{the interaction term is
also written in terms of} $F$ \emph{and not, as usual, in terms of} $A$. In
Sec. 4 we give the comparison with the experiments, particularly with the
Trouton-Noble experiment. It is shown that the approach with geometric 4D
quantities is in a full agreement with the Trouton-Noble experiment. The
explanation for the null result is very simple and natural. Namely in our
approach all quantities are invariant 4D quantities, which means that their
values are the same in the rest frame of the capacitor and in the moving
frame. We have calculated that the torque (as a geometric 4D quantity) is
zero for the stationary capacitor. Then automatically it follows that the
torque is zero for the moving capacitor as well. In the last section, Sec.
5, the discussion and conclussions are presented. \bigskip \bigskip

\noindent \textbf{2.\ THE\ }$F$\ \textbf{FORMULATION\ OF\ ELECTROMAGNETISM
\bigskip \medskip }

\noindent \textbf{2.1. Generally about Geometric Approach to
Electromagnetism\bigskip }

\noindent As mentioned in Sec. 1 the presented formulation of
electromagnetism with the $F$ field exclusively deals with AQs (thus defined
without reference frames) or with the corresponding CBGQs, when some basis
has been introduced. Usualy it is the standard basis that is introduced,
e.g., [1-3]. The generators of the spacetime algebra (the Clifford algebra
generated by Minkowski spacetime) are taken to be four basis vectors $%
\left\{ \gamma _{\mu }\right\} ,$ $\mu =0...3,$ satisfying $\gamma _{\mu
}\cdot \gamma _{\nu }=\eta _{\mu \nu }=diag(+---).$ This basis, the standard
basis $\left\{ \gamma _{\mu }\right\} $, is a right-handed orthonormal frame
of vectors in the Minkowski spacetime $M^{4}$ with $\gamma _{0}$ in the
forward light cone. The $\gamma _{k}$ ($k=1,2,3$) are spacelike vectors. The
$\gamma _{\mu }$ generate by multiplication a complete basis for the
spacetime algebra: $1,\gamma _{\mu },\gamma _{\mu }\wedge \gamma _{\nu
},\gamma _{\mu }\gamma _{5,}\gamma _{5}$ ($2^{4}=16$ independent elements). $%
\gamma _{5}$ is the pseudoscalar for the frame $\left\{ \gamma _{\mu
}\right\} .$ For more details about geometric algebra see, e.g., [1-4], or
short reviews presented in the second paper in [10] and in [11].

We remark that the standard basis corresponds, in fact, to the specific
system of coordinates, i.e., the Einstein system of coordinates, of the
chosen inertial frame of reference. (In the Einstein system of coordinates
the Einstein synchronization [12] of distant clocks and Cartesian space
coordinates $x^{i}$ are used in the chosen inertial frame of reference.)
However different systems of coordinates of an inertial frame of reference
are allowed and they are all equivalent in the description of physical
phenomena. For example, in [13] (and the second and third paper in [14]) two
very different, but completely equivalent systems of coordinates, the
Einstein system of coordinates and ''radio'' (''r'') system of coordinates,
are exposed and exploited throughout the paper.

Any Clifford multivector $A$, an AQ, can be written as a CBGQ, thus with
components and a basis. \emph{Any CBGQ is an invariant quantity upon the
Lorentz transformations (LT).} In such an interpretation the LT are
considered as passive transformations; both the components and the basis
vectors are transformed but the whole 4D geometric quantity remains
unchanged, e.g., the position 1-vector $x$ can be decomposed in the $S$ and $%
S^{\prime }$ (relatively moving) frames and in the standard basis $\left\{
\gamma _{\mu }\right\} $ and some non-standard basis $\left\{ e_{\mu
}\right\} $ as $x=x^{\mu }\gamma _{\mu }=x^{\prime \mu }\gamma _{\mu
}^{\prime }=....=x_{e}^{\prime \mu }e_{\mu }^{\prime }.$ The primed
quantities are the Lorentz transforms of the unprimed ones.

However in the usual Clifford algebra formalism, e.g., [1-4], one deals with
the multivectors as AQs and the LT are considered as active transformations.
If some basis is introduced (for example, the $\left\{ \gamma _{\mu
}\right\} $ basis) then the components of, e.g., some 1-vector relative to a
given inertial frame of reference (with the standard basis $\left\{ \gamma
_{\mu }\right\} $) are transformed into the components of a new 1-vector
relative to the same frame (the basis $\left\{ \gamma _{\mu }\right\} $ is
not changed). (We note that a coordinate-free form for the LT is presented
in [13] and [15] and it can be used both in an active way, when there is no
basis, or in a passive way, when some basis is introduced.) In this paper,
for the sake of brevity and of clearness of the whole exposition, we shall
work either with 4D AQs or with 4D CBGQs which are written only in the
standard basis $\left\{ \gamma _{\mu }\right\} $, but remembering that the
approach with geometric 4D quantities holds for any choice of the
basis.\bigskip \medskip

\noindent \textbf{2.2. The\ Determination of the Electromagnetic Field }$F$
\bigskip

\noindent We start the exposition of the classical electromagnetism by the
description of the first step in the field view of particle-to-particle
interaction; the determination of $F$ for the given sources. As it is
already said this is an axiomatic formulation of the electromagnetism with
only one postulated equation; it is the field equation written in terms of $%
F $ [1-3] (a single field equation for $F$ is first given by M. Riesz [16]).
In that equation an electromagnetic field is represented by a
bivector-valued function $F=F(x)$ on the spacetime. The source of the field
is the electromagnetic current $j$ which is a 1-vector field. Using that the
gradient operator $\partial $ is a 1-vector field this equation can be
written as
\begin{equation}
\partial F=j/\varepsilon _{0}c,\quad \partial \cdot F+\partial \wedge
F=j/\varepsilon _{0}c.  \label{MEF}
\end{equation}
The trivector part is identically zero in the absence of magnetic charge.

When (\ref{MEF}) is written with CBGQs in the $\left\{ \gamma _{\mu
}\right\} $ basis it becomes
\begin{equation}
\partial _{\alpha }F^{\alpha \beta }\gamma _{\beta }-(1/2)\varepsilon
^{\alpha \beta \gamma \delta }\partial _{\alpha }F_{\gamma \delta }\gamma
_{5}\gamma _{\beta }=(1/\varepsilon _{0}c)j^{\beta }\gamma _{\beta },
\label{c1}
\end{equation}
where $\varepsilon ^{\alpha \beta \gamma \delta }$ is the totally
skew-symmetric Levi-Civita pseudotensor. In (\ref{c1}) AQs from (\ref{MEF})
are written as CBGQs in the $\left\{ \gamma _{\mu }\right\} $ basis; $%
F=(1/2)F^{\alpha \beta }\gamma _{\alpha }\wedge \gamma _{\beta }$ (the basis
components $F^{\alpha \beta }$ are determined as $F^{\alpha \beta }=\gamma
^{\beta }\cdot (\gamma ^{\alpha }\cdot F)=(\gamma ^{\beta }\wedge \gamma
^{\alpha })\cdot F$). From (\ref{c1}) one easily finds the usual covariant
form (thus only the basis components of the 4D geometric quantities in the $%
\left\{ \gamma _{\mu }\right\} $ basis) of the field equations as
\begin{equation}
\partial _{\alpha }F^{a\beta }=j^{\beta }/\varepsilon _{0}c,\quad \partial
_{\alpha }\ ^{\ast }F^{\alpha \beta }=0,  \label{maxco}
\end{equation}
where the usual dual tensor is introduced $^{\ast }F^{\alpha \beta
}=(1/2)\varepsilon ^{\alpha \beta \gamma \delta }F_{\gamma \delta }$.

The field bivector $F$ yields the complete description of the
electromagnetic field and, in fact, there is no need to introduce either the
field vectors or the potentials. For the given sources the Clifford algebra
formalism enables one to find in a simple way the electromagnetic field $F.$
Namely the gradient operator $\partial $ is invertible and (\ref{MEF}) can
be solved for
\begin{equation}
F=\partial ^{-1}(j/\varepsilon _{0}c),  \label{inef}
\end{equation}
see, e.g., [17] and [1] Spacetime Calculus. We briefly repeat the main
points related to (\ref{inef}) from these references. However, the important
difference with respect to the usual approaches [1-3] is that for us, as
proved in [10] and [11], the field equation (\ref{MEF}) is not equivalent to
the usual Maxwell equations with the 3D $\mathbf{E}$ and $\mathbf{B}$ (i.e.,
with $\mathbf{E}_{H},$ $\mathbf{B}_{H}$ from (\ref{hf}) or $\mathbf{E}_{J},$
$\mathbf{B}_{J}$ from (\ref{J1})). $\partial ^{-1}$ is an integral operator
which depends on boundary conditions on $F$ and (\ref{inef}) is an integral
form of the field equation (\ref{MEF}). If the charge-current density $j(x)$
is the sole source of $F,$ then (\ref{inef}) provides the unique solution to
the field equation (\ref{MEF}). By using Gauss' Theorem an important formula
can be found that allows to calculate $F$ at any point $y$ inside
m-dimensional manifold $M$ from its derivative $\partial F$ and its values
on the boundary $\partial M$ if a Green's function $G(y,x)$ is known,
\begin{equation}
F(y)=\int_{M}G(y,x)\partial F(x)\mid d^{m}x\mid -\int_{\partial
M}G(y,x)n^{-1}F(x)\mid d^{m-1}x\mid ,  \label{DF}
\end{equation}
$n$ is a unit normal, $n^{-1}=n$ if $n^{2}=1$ or $n^{-1}=-n$ if $n^{2}=-1,$
and $G(y,x)$ is a solution to the differential equation $\partial
_{y}G(y,x)=\delta ^{m}(y-x).$ ((\ref{DF}) is the relation (4.17) in [17].)
If $\partial F=0$, i.e., $j=0$, the first term on the right side of (\ref{DF}%
) vanishes but not the second term. This general relation can be applied to
different examples.\bigskip \medskip

\noindent \textbf{2.3. The\ Electromagnetic Field of a Point Charge\bigskip }

\noindent An example is the determination of the expression for the
classical Li\'{e}nard-Wiechert field that is given, e.g., in [17] and [1]
Spacetime Calculus. The usual procedure ([17] and [1]) is to utilize the
general relation (\ref{DF}), in which all quantities are defined without
reference frames (Geometric calculus), and to specify it to the Minkowski
spacetime ($m=4$). Then a space-time split is introduced by the relation
\begin{equation}
ct=x\cdot n=x\cdot \gamma _{0}.  \label{sp2}
\end{equation}
($n$ in (\ref{DF}) is taken to be $\gamma _{0}$ and (\ref{sp2}) is the
equation for a 1-parameter family of spacelike hyperplanes $S(t)$ with
normal $\gamma _{0}$; $S(t)$ is a surface of simultaneous $t$ when the
Einstein synchronization is chosen.) For simplicity, $M$ is taken to be the
entire region between the hyperplanes $S_{1}=S(t_{1})$ and $S_{2}=S(t_{2}).$
We shall not discuss this derivation further but we only quote the result
for the classical Li\'{e}nard-Wiechert field. The charge-current density for
a particle with charge $q$ and world line $z=z(\tau )$ with proper time $%
\tau $ is $j(x)=q\int_{-\infty }^{\infty }d\tau u\delta ^{4}(x-z(\tau )),$
where $u=u(\tau )=dz/d\tau $. Then the classical Li\'{e}nard-Wiechert
retarded field for $q$ (see, e.g., Sec. 5 in [17]) is
\begin{equation}
F(x)=(q/4\pi \varepsilon _{0})\{r\wedge \lbrack (u/c)+(1/c^{3})r\cdot
(u\wedge \overset{\cdot }{u})]\}/(r\cdot u/c)^{3},  \label{LW}
\end{equation}
where $r=x-z$ satisfies the light-cone condition $r^{2}=0$ and $z,$ $u,$ $%
\overset{\cdot }{u}=du/d\tau $ are all evaluated at the intersection of the
backward light cone (with vertex at $x$) and world line of that charge $q$.
It is worth noting that from the general expression (\ref{DF}) one can
derive not only the retarded interpretation for $F$ of a charge $q$ but also
the advanced interpretation and the present-time interpretation, i.e., an
instantaneous action-at-a-distance interpretation. (This present-time
interpretation will be reported elsewhere. In the tensor formalism the
expressions for $F^{ab}$ and the 4-vectors $E^{a}$ and $B^{a}$ in the
present-time interpretation for an uniform and uniformly accelerated motion
of a charge $q$ are given in [18].)

All quantities in (\ref{LW}) are geometric 4D quantities, the AQs, and for
more practical use they can be written as CBGQs, usually in the $\left\{
\gamma _{\mu }\right\} $ basis. Thus, the general expression for $F$ for an
arbitrary motion of a charge is
\begin{align}
F& =(1/2)F^{\alpha \beta }\gamma _{\alpha }\wedge \gamma _{\beta },\quad
F^{\alpha \beta }=(kq/\zeta ^{3})\left[ c^{2}(r^{\alpha }u^{\beta }-r^{\beta
}u^{\alpha })\right]  \notag \\
& +(kq/\zeta ^{3})\left[ (r^{\sigma }\overset{\cdot }{u}_{\sigma
})(r^{\alpha }u^{\beta }-r^{\beta }u^{\alpha })+(r^{\sigma }u_{\sigma
})(r^{\alpha }\overset{\cdot }{u}^{\beta }-r^{\beta }\overset{\cdot }{u}%
^{\alpha })\right] .  \label{efa}
\end{align}
In (\ref{efa}), $r^{\mu }=x^{\mu }-z^{\mu }(\tau ),$ $x^{\mu }$ and $z^{\mu
}(\tau )$ are the field and the source basis components of $x$ and $z$
respectively (in the $\left\{ \gamma _{\mu }\right\} $ basis), $k=1/4\pi
\varepsilon _{0}$ and $\zeta \equiv r^{\sigma }u_{\sigma }$. The right-hand
side has to be evaluated at $\tau _{0}$ such that $x^{\mu }-z^{\mu }(\tau
_{0})$ is light-like, i.e., $\tau _{0}$ is determined by the above mentioned
light-cone condition, which in the component form becomes $(x^{\sigma
}-z^{\sigma }(\tau _{0}))(x_{\sigma }-z_{\sigma }(\tau _{0}))=0,$ and it
holds that $x^{0}-z^{0}(\tau _{0})=\mid r\mid \succ 0.$ The expression for $%
F^{\alpha \beta }$ from (\ref{efa}) is the standard result, e.g., Jackson's
book [5]. The first term in $F^{\alpha \beta }$ (\ref{efa}) represents the
velocity part and the second one represents the acceleration or radiation
part. (However we note that such decomposition of $F^{\alpha \beta }$ into
velocity and acceleration parts is the consequence of the used retarded
representation and it does not exist for, e.g., the present time
representation [18].) We see that from (\ref{LW}) one can simply find the
well-known result for $F^{\alpha \beta }$ given in (\ref{efa}).

The components $F^{\alpha \beta }$ are measurable quantities; they can be
measured using the Lorentz force and Newton's second law as will be
discussed in Sec. 2.5.

Let us specify the above relations to the case of a point charge with \emph{%
constant velocity} $u$, see for the comparison Sec. 7.3.2 in [2]\textit{.}
Then the trajectory is $z(\tau )=u\tau $ (taking that $z(0)=0$), $r\cdot
u/c=\left| x\wedge (u/c)\right| $, $r\wedge (u/c)=x\wedge (u/c)$.
Substituting these relations into (\ref{LW}) one finds the field strength $F$
\begin{equation}
F(x)=kq(x\wedge (u/c))/\left| x\wedge (u/c)\right| ^{3}=D(x\wedge (u/c)),
\label{cvf}
\end{equation}
where $D=kq/\left| x\wedge (u/c)\right| ^{3}$.

In all usual formulations of electromagnetism both in the Clifford algebra
[1-3] and tensor formalisms [5] the results (\ref{LW}) or $F^{\alpha \beta }$
from (\ref{efa}) are considered only as \emph{formal, mathematical results}
that are necessary to find the components of the ``physical'' quantities,
the 3D $\mathbf{E}$ and $\mathbf{B}$. But the relations (\ref{inef}), (\ref
{DF}) and (\ref{LW}) show that $F$ has an independent physical reality and
the whole electromagnetism can be treated with $F$ without even mentioning
the 3D $\mathbf{E}$ and $\mathbf{B}$. Consequently, in contrast to the usual
approaches [1-3], and all other previous approaches, we assume that the 4D
geometric quantity $F$ \emph{can be considered as the primary physical
quantity} and not the 3D vectors $\mathbf{E}$ and $\mathbf{B}$. Then from
the known $F$ one can find different 4D quantities that represent the 4D
electric and magnetic fields; they are considered in [10,11] and [15].

One of these representations uses the decomposition of $F$ into 1-vectors $E$
and $B$
\begin{align}
F& =(1/c)E\wedge v+(IB)\cdot v,  \notag \\
E& =(1/c)F\cdot v,\quad B=-(1/c^{2})I(F\wedge v),  \label{itf}
\end{align}
where $I$ is the unit pseudoscalar. ($I$ is defined algebraically without
introducing any reference frame, as in [4] Sec. 1.2.) It holds that $E\cdot
v=B\cdot v=0$ (since $F$ is skew-symmetric). $v$ in (\ref{itf}) can be
interpreted as the velocity (1-vector) of a family of observers who measures
$E$ and $B$ fields. \emph{The velocity} $v$ \emph{and all other quantities
entering into} (\ref{itf}) \emph{are defined without reference frames; they
are AQs.} $v$ characterizes some general observer. Thus \emph{the relations }%
(\ref{itf})\emph{\ hold for any observer.} (For the equivalent relations
with $v$ in the tensor formalism see also [19].) The relations (\ref{itf})
actually establish the equivalence of the formulation of electromagnetism
with the field bivector $F$ (presented here) and the formulation with
1-vectors of the electric $E$ and magnetic $B$ fields (that is presented in
[15]). \emph{Both formulations, with} $F$ \emph{and} $E,$ $B$ \emph{fields,
are equivalent formulations, but every of them is a complete, consistent and
self-contained formulation.}

Similarly in [11] (see also the second paper in [10]) it is shown that the
relations (\ref{hf}) from [1,2] have to be replaced by the relations
\begin{align}
F& =E_{Hv}+cIB_{Hv}\mathbf{,\quad }E_{Hv}=(1/c^{2})(F\cdot v)\wedge v  \notag
\\
B_{Hv}& =-(1/c^{3})I[(F\wedge v)\cdot v].  \label{he}
\end{align}
Namely in (\ref{hf}) a space-time split in the $\gamma _{0}$ - frame is used
and, as it is already said, it depends on the velocity $c\gamma _{0}$ of the
specific observer, the $\gamma _{0}$ - observer. Thus in (\ref{hf}) the
\emph{observer independent} quantity $F$ is decomposed into the \emph{%
observer dependent} quantities $\mathbf{E}_{H}$, $\mathbf{B}_{H}$ defined
only in the $\gamma _{0}$ - frame. A space-time split is not a Lorentz
invariant procedure. On the other hand in (\ref{he}) the velocity $c\gamma
_{0}$ of the specific observer is replaced by the velocity $v$ of some
general observer. $v$ is an AQ and thence the relations (\ref{he}), in the
same way as the relations (\ref{itf}), hold for any observer. Then, instead
of formulating the whole electromagnetism by 1-vectors of the electric and
magnetic fields, $E$ and $B$ respectively, one can formulate it by bivectors
$E_{Hv}$ and $B_{Hv}$. The only difference is that the decomposition of $F$
into 1-vectors $E$ and $B$ (\ref{itf}) is much simpler and, in fact, closer
to the classical formulation of the electromagnetism with the 3D $\mathbf{E}$
and $\mathbf{B}$, than the decomposition of $F$ into bivectors $E_{Hv}$ and $%
B_{Hv}$ (\ref{he}). In contrast to the formulation with $F$, all
formulations with the electric and magnetic fields as AQs require the
introduction of $v$, the velocity of observers who measure fields. This is
the reason why the formulation with the bivector field $F$ is investigated
and presented in this paper.

As an example let us apply Eqs. (\ref{cvf}) and (\ref{he}) to determine the
electric and magnetic fields $E_{Hv}$ and $B_{Hv}$ for the case of a point
charge with constant velocity $u$. We find
\begin{eqnarray}
E_{Hv} &=&(D/c^{3})[(u\cdot v)(x\wedge v)-(x\cdot v)(u\wedge v)]  \notag \\
B_{Hv} &=&(D/c^{4})I[(x\cdot v)(u\wedge v)-(u\cdot v)(x\wedge
v)+c^{2}(x\wedge u)].  \label{bi}
\end{eqnarray}
Both $E_{Hv}$ and $B_{Hv}$ from (\ref{bi}) are AQs, i.e., they are defined
without reference frames. Remember that $v$ is the velocity (1-vector) of a
family of observers who measures $E_{Hv}$ and $B_{Hv}$ fields and $u$ is the
velocity (1-vector) of a point charge.

Using (\ref{itf}) and $F$ from (\ref{cvf}) we get physically equivalent but
simpler expressions
\begin{eqnarray}
E &=&(D/c^{2})[(u\cdot v)x-(x\cdot v)u]  \notag \\
B &=&(-D/c^{3})I(x\wedge u\wedge v),  \label{ec}
\end{eqnarray}
in which the electric and magnetic fields are represented by 1-vectors $E$
and $B$.

We note that ($E$, $B$) from (\ref{ec}) or $E_{Hv}$, $B_{Hv}$ from (\ref{bi}%
) and $F$ from (\ref{cvf}) contain the same amount of physical informations.
The expressions with AQs ($E$, $B$) from (\ref{ec}) or $E_{Hv}$, $B_{Hv}$
from (\ref{bi}) were not found in any previous approach including [1-3]. The
usual results can be recovered simply taking that the observers who measure $%
E$, $B$ or $E_{Hv}$, $B_{Hv}$ fields are at rest, ``fiducial'' observers
(the $\gamma _{0}$-frame with the $\left\{ \gamma _{\mu }\right\} $ basis),
for which $v=c\gamma _{0}$ in (\ref{bi}) ((\ref{ec})). In that case, e.g.,
the relations (\ref{ec}) become
\begin{equation}
E_{f}=D(\gamma x-ut),\quad B_{f}=(-D/c^{2})\gamma _{5}(x\wedge u\wedge
\gamma _{0}),  \label{fe}
\end{equation}
where $f$ stands for ``fiducial'' and $\gamma _{5}$ is the unit pseudoscalar
$I$ for the $\left\{ \gamma _{\mu }\right\} $ basis. Notice that $%
E_{f}^{0}=B_{f}^{0}=0$, which means that in the frame of ``fiducial''
observers, the $\gamma _{0}$-frame with the $\left\{ \gamma _{\mu }\right\} $
basis, $E_{f}$, $B_{f}$ contain only spatial components. Thence $E_{f}$ and $%
B_{f}$ from (\ref{fe}) are exactly the same as the usual expressions for the
3D electric and magnetic fields of a charge in uniform motion. \textbf{%
\bigskip \medskip }

\noindent \textbf{2.4. The Integral Form of the Field Equation }(\ref{MEF}%
)\bigskip

\noindent Instead of dealing with the axiomatic formulation of
electromagnetism that uses only the \emph{local }form of the field equation%
\textbf{\ }(\ref{MEF}) one can construct the equivalent integral form. Such
form is constructed by Hestenes and nicely presented in [17] and [1]
Space-Time Calculus, though, Hestenes does not consider it as an axiomatic
formulation. Here only the main results from [17] and [1] will be briefly
repeated and applied to the determination of $F$ (and $E$, $B$) for some
simple cases.

The trivector part of (\ref{MEF}) $\partial \wedge F=0$ can be transformed
to an equivalent integral form as
\begin{equation}
\oint_{\partial M}d^{2}x\cdot F=0,  \label{d0}
\end{equation}
where in the directed integral (\ref{d0}), in general, $d^{m}x$ is the
directed measure and $\partial M$ is any closed 2-dimensional submanifold in
spacetime. $d^{m}x$ can be resolved into its magnitude $\left| d^{m}x\right|
$, which is the usual ``scalar measure,'' and its direction represented by a
unit $m$-blade $I_{m}$: $d^{m}x=I_{m}\left| d^{m}x\right| $, or in terms of
CBGQs $d^{m}x=d_{1}x\wedge d_{2}x\wedge ...\wedge d_{m}x=e_{1}\wedge
e_{2}\wedge ...e_{m}dx^{1}dx^{2}...dx^{m}$, where $dx^{\mu }$ is a scalar
differential for the scalar variable $x^{\mu }$ and $e_{\mu }$ is the basis
1-vector for some basis $\left\{ e_{\mu }\right\} $.

In order to find the corresponding integral forms with the electric and
magnetic fields Hestenes, [17] and [1], uses the space-time split given by (%
\ref{hf}) and shows that (\ref{d0}) is equivalent to Faraday's law or ``the
absence of magnetic poles,'' or a mixture of the two, depending on the
choice of $\partial M$\textit{.} However, as we have said, the use of such
procedure, a space-time split in the $\gamma _{0}$-frame, transforms the
integral field equation (\ref{d0}) \emph{written in terms of AQs} into \emph{%
observer dependent} integral field equation written in terms of \emph{%
observer dependent }$\mathbf{E}_{H}$ \emph{and} $\mathbf{B}_{H}$ from (\ref
{hf}).

Instead of using such procedure we express $F$ in (\ref{d0}) in terms of
\emph{AQs} $E_{Hv}$\textit{, }$B_{Hv}$ from (\ref{he}) or $E$, $B$ from (\ref
{itf}) to find equivalent, coordinate-free, integral forms with electric and
magnetic fields as AQs, e.g., with $E$ and $B$%
\begin{equation}
\oint_{\partial M}d^{2}x\cdot ((1/c)E\wedge v+(IB)\cdot v)=0.  \label{db}
\end{equation}
Then going to the frame of ``fiducial'' observers (for which $v=c\gamma _{0}$%
), the $\gamma _{0}$-frame with the $\left\{ \gamma _{\mu }\right\} $ basis,
$E$ and $B$ contain only spatial components, and we recover the usual
integral form of Faraday's law and Gauss' law for a magnetic field or a
mixture of the two, depending on the choice of $\partial M$.

Hestenes also derived an integral formula for the vector part of the local
field equation\textbf{\ }(\ref{MEF})
\begin{equation}
\oint_{\partial M}d^{2}x\cdot (FI)=(1/\varepsilon _{0}c)\int_{M}j\cdot
n\left| d^{3}x\right| ,  \label{dt}
\end{equation}
where $n=n(x)$ is a unit outward normal and $M$ is any 3-dimensional
submanifold in spacetime that is enclosed by $\partial M$. In [17] and [1]
the equation (\ref{dt}) is also written in the less familiar form
\begin{equation}
\oint_{\partial M}d^{2}x\wedge F=(1/\varepsilon _{0}c)\int_{M}(d^{3}x)\wedge
j,  \label{ds}
\end{equation}
which when combined with (\ref{d0}) gives the integral version of the local
field equation\textbf{\ }(\ref{MEF})
\begin{equation}
\oint_{\partial M}\left\langle d^{2}xF\right\rangle _{I}=(1/\varepsilon
_{0}c)\int_{M}\left\langle d^{3}xj\right\rangle _{I},  \label{dk}
\end{equation}
where $\left\langle ...\right\rangle _{I}$ selects only the ``invariant (=
scalar+pseudoscalar) parts.'' Of course the whole discussion that led from (%
\ref{d0}) to (\ref{db}) applies in the same measure to the equations (\ref
{dt}) (or (\ref{ds})) and (\ref{dk}) and the corresponding expressions with
\emph{AQs} $E_{Hv}$\textit{, }$B_{Hv}$ from (\ref{he}) or $E$, $B$ from (\ref
{itf}). We notice that when $F$ is decomposed in terms of electric and
magnetic fields then the equation (\ref{dt}), or (\ref{ds}), in the frame of
``fiducial'' observers ($v=c\gamma _{0}$) becomes the Amp\`{e}re-Maxwell
law, the Gauss law for an electric field or a mixture of the two, depending
on the choice of $\partial M$.

In the usual approach with the 3D $\mathbf{E}$ and $\mathbf{B}$ and for some
simple cases one can determine, e.g., the 3D $\mathbf{E}$, directly from the
integral version of the Gauss law. Similarly instead of to find $F$ from (%
\ref{DF}) one can get it, e.g., from the equation (\ref{dt}) (or (\ref{ds}%
)). Let us consider a flat sheet infinite in extent, with the constant
surface charge density $\sigma $. We can use (\ref{itf}) to connect the $F$
formulation with the formulation that deals with the electric and magnetic
fields $E$ and $B$ respectively. For the sake of easier comparison with the
common approach that deals with the 3D $\mathbf{E}$ and $\mathbf{B}$ we
introduce the $\gamma _{0}$-frame (with the $\left\{ \gamma _{\mu }\right\} $
basis) in which the flat sheet is at rest and situated in the $\gamma
_{1}\wedge \gamma _{2}$ plane. In the usual application of Gauss' law with
the 3D $\mathbf{E}$ a convenient Gaussian surface in the 3D space is chosen
to be a ``pill box'' piercing a cross-sectional area $S$ on the flat sheet
and whose height is $2a$. In the 4D spacetime $\partial M$ in the equation (%
\ref{dt}) is chosen to be the same closed 2-dimensional surface as above (a
``pill box'') that is \emph{instantaneously} taken, i.e., it is at rest, in
the $\gamma _{0}$-frame. When $F$ is written as a CBGQ in the $\left\{
\gamma _{\mu }\right\} $ basis it becomes
\begin{equation}
F=F^{i0}\gamma _{i}\wedge \gamma _{0}+(1/2)F^{kl}\gamma _{k}\wedge \gamma
_{l}.  \label{fi}
\end{equation}
This is a decomposition of $F$ into ``electriclike'' (the first part) and
``magneticlike'' (the second part), but we note that such decomposition of $%
F $ as in (\ref{fi}) is not a Lorentz invariant decomposition. From (\ref
{itf}), i.e., from (\ref{fi}) (the chosen $\gamma _{0}$-frame is the frame
of ``fiducial'' observers, $v=c\gamma _{0}$), and by analogy with the
corresponding 3D formulation, we conclude that only $F^{30}\neq 0$ and it is
of constant magnitude. Then the equation (\ref{dt}) becomes $%
2SF^{30}=S(\sigma /\varepsilon _{0})$ whence
\begin{equation}
F=F^{30}\gamma _{3}\wedge \gamma _{0}=(\sigma /2\varepsilon _{0})\gamma
_{3}\wedge \gamma _{0}  \label{fp}
\end{equation}
This example will be used in comparison with experiments that is considered
in Sec. 4.

In our axiomatic formulation of electromagnetism there is \emph{only one
postulated equation}, either the local field equation\textbf{\ }(\ref{MEF})
or its equivalent integral version (\ref{dk}), whereas, as we have already
mentioned, in the recent axiomatic formulation of electromagnetism [8] \emph{%
there are three postulated equations}; (1) electric charge conservation, (2)
the Lorentz force, (3) magnetic flux conservation. Using these three
postulated equations and \emph{foliation of spacetime} (what is nothing else
than the space-time split) the authors of [8] derive the usual form of the
Maxwell equations with the 3D $\mathbf{E}$ and $\mathbf{B}$. Their axiom 3
simply follows from the equation (\ref{d0}) and in the next sections, Secs.
2. 6 and 2. 5, we shall show that not only axiom 3 but also the axioms 1 and
2 from [8] simply follow from our equation (\ref{MEF}), or (\ref{dk}).
\medskip \bigskip

\noindent \textbf{2.5. The\ Lorentz Force and the Motion of a Charged }

\noindent \textbf{Particle in the Electromagnetic Field }$F$ \bigskip

\noindent As it is said in the field view of particle-to-particle
interaction the electrodynamic interaction between charges is described as
two-steps process; first fields are seen as being generated from their
particle sources and then the fields so generated are perceived as
interacting with some target particle. The description of the first step in
the $F$ formulation of electrodynamics is given by the above relations (\ref
{MEF}) (or (\ref{dk})), (\ref{inef}), (\ref{DF}) and for a point particle
with charge $q$ with (\ref{LW}). The second step requires the determination
of the Lorentz force in terms of $F$ and its use in Newton's second law.
This will be undertaken below.

In the Clifford algebra formalism one can easily derive the expressions for
the stress-energy vector $T(n)$ and the Lorentz force density $K_{(j)}$
directly from the field equation (\ref{MEF}) and from the equation for $%
\widetilde{F},$ the reverse of $F,$ $\widetilde{F}\widetilde{\partial }=%
\widetilde{j}/\varepsilon _{0}c$ ($\widetilde{\partial }$ differentiates to
the left instead of to the right). Indeed, using (\ref{MEF}) and from the
equation for $\widetilde{F}$ one finds
\begin{equation}
T(\partial )=(-\varepsilon _{0}/2)(F\partial F)=j\cdot F/c=-K_{_{(j)}},
\label{TEF}
\end{equation}
where in $(F\partial F)$ the derivative $\partial $ operates to the left and
to the right by the chain rule. The stress-energy vector $T(n)$ [1-3] for
the electromagnetic field is then defined in the $F$ formulation as
\begin{equation}
T(n)=T(n(x),x)=-(\varepsilon _{0}/2)\left\langle FnF\right\rangle _{1}.
\label{ten}
\end{equation}
We note that $T(n)$ is a vector-valued linear function on the tangent space
at each spacetime point $x$ describing the flow of energy-momentum through a
hypersurface with normal $n=n(x)$.

The right hand side of (\ref{TEF}) yields the expression for the Lorentz
force density $K_{(j)},$
\begin{equation}
K_{(j)}=F\cdot j/c.  \label{lf}
\end{equation}
The Lorentz force in the $F$ formulation for a charge $q$ is $K=(q/c)F\cdot
u $, where $u$ is the velocity 1-vector of a charge $q$ (it is defined to be
the tangent to its world line).

It is worth noting that the equation (\ref{lf}) is the second postulate,
axiom 2, in the axiomatic formulation of electromagnetism [8]. Thus in our
approach Eq. (\ref{lf}) simply follows from the field equation (\ref{MEF}).

In the approaches [1,2] the Lorentz force is discussed using the space-time
split and the corresponding decomposition of $F$ into the electric and
magnetic components. However in the analysis of the motion of a charged
particle under the action of the Lorentz force we utilize only those parts
of the usual approaches [1-3] that are expressed only in terms of $F$ and
not those expressed by $\mathbf{E}_{H}$, $\mathbf{B}_{H}$\textbf{\ }or $%
\mathbf{E}_{J},$ $\mathbf{B}_{J}.$ We shall only quote the main results from
[1-3] for the motion of a charged particle in a constant electromagnetic
field $F$ but without using $\mathbf{E}_{H}$, $\mathbf{B}_{H}$\textbf{\ }or $%
\mathbf{E}_{J},$ $\mathbf{B}_{J}$. Actually, as shown in Sec. 2.3, instead
of dealing with the decomposition of $F$ as an \emph{AQ} into the \emph{%
observer dependent} $\mathbf{E}_{H}$, $\mathbf{B}_{H}$\textbf{\ }(\ref{hf})
or $\mathbf{E}_{J},$ $\mathbf{B}_{J}$ (\ref{J1}) as in [1-3] one has to make
the decomposition of the \emph{AQ} $F$ into \emph{AQs} $E_{Hv}$, $B_{Hv}$ (%
\ref{he}) \emph{or} $E$, $B$ (\ref{ec}). Then going to the frame of
``fiducial'' observers, $v=c\gamma _{0}$, one recovers \emph{in that frame}
the usual results [1-3] with the electric and magnetic fields.

The particle equation of motion, i.e., Newton's second law is
\begin{equation}
m\overset{\bullet }{u\ }=qF\cdot u,  \label{LNF}
\end{equation}
where $\overset{\bullet }{u}\ =du/d\tau ;$ the overdot denotes
differentation with respect to proper time $\tau $. Usually [1,2] the
equation (\ref{LNF}) is not solved directly but solving the rotor equation $%
\overset{\bullet }{R}\ =(q/2m)FR$ and using the invariant canonical form for
$F,$ which is
\begin{equation}
F=fe^{I\varphi }=f(\cos \varphi +I\sin \varphi );  \label{fc}
\end{equation}
this holds for $F^{2}\neq 0$. In that form $f^{2}=\left| f^{2}\right| $,
which shows that $f$ is a ``timelike bivector,'' but $If$ is a ``spacelike
bivector,'' since $(If)^{2}=-\left| f^{2}\right| $. The equation (\ref{fc})
is the unique decomposition of $F$ into a sum of mutually commuting timelike
and spacelike parts. As shown in, e.g., [1] Space-Time Calculus, both $f$
and $\varphi $ can be written in terms of $F$, i.e., invariants under the LT
that are constructed from $F$; $\alpha =F\cdot F$, $I\beta =F\wedge F$. Thus
$e^{I\varphi }=(\alpha +I\beta )^{1/2}/(\alpha ^{2}+\beta ^{2})^{1/4}$ and $%
f=F(\alpha -I\beta )^{1/2}/(\alpha ^{2}+\beta ^{2})^{1/4}$. The same
invariant decomposition of $F$ is given in [3] Chap. 6 par.3, where $F$ is
written as $F=F_{1}+F_{2}$ and it can be shown that $F_{1}$ ($F_{2}$) from
[3] is exactly equal to $f\cos \varphi $ ($fI\sin \varphi $) from [1]. Note
the difference between the decomposition of $F$ presented in (\ref{fi}) and
that one given in (\ref{fc}); the first one is a coordinate-dependent
presentation whereas the second one is given in terms of AQs.

Let us consider that $F$ is an uniform electromagnetic field and let us
apply the decomposition (\ref{fc}). Then denoting ($q/m)F=\Omega $, $\Omega
_{1}=f(q/m)\cos \varphi $, $\Omega _{2}=f(q/m)\sin \varphi $, and making an
invariant decomposition of the initial velocity $u(0)$ into a component $%
u_{1}$ in the $f$-plane and a component $u_{2}$ orthogonal to the $f$-plane,
$u(0)=f^{-1}(f\cdot u(0))+f^{-1}(f\wedge u(0))=u_{1}+u_{2}$, we get
\begin{equation}
u=e^{(1/2)\Omega _{1}\tau }u_{1}+e^{(1/2)\Omega _{2}\tau }u_{2}.  \label{veF}
\end{equation}
As stated in [1], Spacetime Calculus, this is an invariant decomposition of
the motion into ''electriclike'' and ''magneticlike'' components. The
particle history is obtained integrating (\ref{veF})
\begin{equation}
x(\tau )-x(0)=(e^{(1/2)\Omega _{1}\tau }-1)\Omega
_{1}^{-1}u_{1}+e^{(1/2)\Omega _{2}\tau }\Omega _{2}^{-1}u_{2}.  \label{iS}
\end{equation}
(For more details see [1-3].) This result applies for arbitrary initial
conditions and arbitrary uniform electromagnetic field $F$. Different
special cases of the equation (\ref{iS}) that correspond to the motion of a
charge in uniform electric or magnetic fields are already considered, using
only $F$, in, e.g., [3], and will not be considered here. Of course all
other special cases, e.g., a charge in an electromagnetic plane wave, can
also be investigated exclusively in terms of $F$ without introducing the
electric and magnetic fields. The solutions for the motion of a charged
particle in a constant electromagnetic field that are similar to (\ref{veF})
and (\ref{iS}) are already considered in the usual covariant approach (thus
with $F^{\mu \nu }$ and not with AQ $F$) in [20].

It is already mentioned that the expression for the Lorentz force in terms
of $F$ determines the way in which $F$, i.e., the components of $F$ in some
reference frame are measured. First let us assume that in the chosen
reference frame with the $\left\{ \gamma _{\mu }\right\} $ basis the
considered charge is at rest, $u=c\gamma _{0}$ (in components $u_{\mu
}=(c,0,0,0)$). Then from the expression for the Lorentz force $K=(q/c)F\cdot
u$ and the decomposition (\ref{fi}) (that holds in the $\left\{ \gamma _{\mu
}\right\} $ basis) we find that only ``electriclike'' part of $F$ is
relevant in that case
\begin{equation}
K_{u_{i}=0}=qF^{i0}\gamma _{i}.  \label{elf2}
\end{equation}
Thence we see that the relation
\begin{equation}
F^{i0}\equiv \lim_{q\rightarrow 0}K_{u_{i}=0}^{i}/q  \label{elf}
\end{equation}
defines experimentally the components (in the $\left\{ \gamma _{\mu
}\right\} $ basis) of ``electriclike'' part of $F$ as the ratio of the
measured force $K_{u_{i}=0}$ on a stationary charge to the charge in the
limit when the charge goes to zero. Having $F^{i0}\gamma _{i}$ so defined
the charge can be given a convenient uniform velocity $u$ with $u_{k}\neq 0$%
, from which the components (in the $\left\{ \gamma _{\mu }\right\} $ basis)
of ``magneticlike'' part of $F$ in the decomposition (\ref{fi}) are defined
from the limit
\begin{equation}
F^{ik}u_{k}\equiv \lim_{q\rightarrow 0}K^{i}/q,\quad i\neq k.  \label{bef}
\end{equation}
It is worth noting that in the usual approaches the components of the 3D $%
\mathbf{E}$ and $\mathbf{B}$ are defined experimentally in exactly the same
way through the 3D Lorentz force.

Instead of such coordinate-dependent formulation we can generalize relations
(\ref{elf}) and (\ref{bef}) comparing them with the results for measuring $E$
and $B$ from [15]. Let us introduce as in (\ref{itf}) the velocity 1-vector $%
v$ of a family of observers who measures $F$ field and consider a special
case, the Lorentz force acting on a charge as measured by a comoving
observer ($v=u$). Then from the definition of $K$ and (\ref{itf}) one finds
that $K=(q/c)F\cdot u=(q/c)F\cdot v=qE$. Thence we can say that the Lorentz
force ascribed by an observer comoving with a charge is \emph{purely electric%
} and we define
\begin{equation}
F_{E}\cdot v/c\equiv \lim_{q\rightarrow 0}K_{v=u}/q,  \label{fv}
\end{equation}
where $F_{E}$ is ``electriclike'' part of $F$. In the $\gamma _{0}$-frame
with the $\left\{ \gamma _{\mu }\right\} $ basis (\ref{fv}) reduces to (\ref
{elf2}) and (\ref{elf}) since $v=c\gamma _{0}$. Having $F_{E}$ so defined
the charge can be given a convenient uniform velocity $u\neq v$ from which
``magneticlike'' part of $F$ can be defined from the limit
\begin{equation}
F\cdot u/c\equiv \lim_{q\rightarrow 0}K/q.  \label{bl}
\end{equation}
In the $\gamma _{0}$-frame with the $\left\{ \gamma _{\mu }\right\} $ basis
and when the definitions (\ref{elf2}) (or (\ref{elf})) are also used then
the relation (\ref{bl}) reduces to (\ref{bef}). This completely defines the
manner in which $F$ is measured by an arbitrary observer. All this together
explicitly shows that $F$ \emph{is a measurable quantity }with a
well-defined physical procedure for the measurement. \medskip \bigskip

\noindent \textbf{2.6. The} \textbf{Stress-Energy Vector} $T(n)$ \textbf{and
the Quantities}

\noindent \textbf{Derived from} $T(n)$\textbf{\bigskip }

\noindent The most important quantity for the momentum and energy of the
electromagnetic field is the observer independent stress-energy vector $T(n)$
(\ref{ten}). It can be written in the following form
\begin{equation}
T(n)=-(\varepsilon _{0}/2)\left[ (F\cdot F)n+2(F\cdot n)\cdot F\right] .
\label{ten1}
\end{equation}
We present a \emph{new form} for $T(n)$ (\ref{ten1}) writing it as a sum of $%
n$-parallel part ($n-\parallel $) and $n$-orthogonal part ($n-\perp $)
\begin{align}
T(n)& =-(\varepsilon _{0}/2)\left[ (F\cdot F)+2(F\cdot n)^{2}\right] n
\notag \\
& -\varepsilon _{0}\left[ (F\cdot n)\cdot F-(F\cdot n)^{2}n\right] .
\label{ste}
\end{align}
The first term in (\ref{ste}) is $n-\parallel $ part and it yields the
energy density $U.$ Namely using $T(n)$ and the fact that $n\cdot T(n)$ is
positive for any timelike vector $n$ we construct the expression for \emph{%
the observer independent energy density} $U$ contained in an electromagnetic
field as $U=n\cdot T(n)=\left\langle nT(n)\right\rangle ,$ (scalar, i.e.,
grade-0 multivector). Thus in terms of $F$ and (\ref{ste}) $U$ becomes
\begin{equation}
U=(-\varepsilon _{0}/2)\left\langle FnFn\right\rangle =-(\varepsilon _{0}/2)
\left[ (F\cdot F)+2(F\cdot n)^{2}\right] .  \label{uen1}
\end{equation}
The second term in (\ref{ste}) is $n-\perp $ part and it is $(1/c)S$, where $%
S$ is \emph{the observer independent expression for the Poynting vector }%
(1-vector),
\begin{equation}
S=-\varepsilon _{0}c\left[ (F\cdot n)\cdot F-(F\cdot n)^{2}n\right] ,
\label{po1}
\end{equation}
and, as can be seen, $n\cdot S=0$. Thus $T(n)$ expressed by $U$ and $S$ is
\begin{equation}
T(n)=Un+(1/c)S.  \label{uis}
\end{equation}
Notice that the decompositions of $T(n)$, (\ref{ten1}), (\ref{ste}) and (\ref
{uis}), are all observer independent decompositions, thus with AQs. Further
\emph{the observer independent momentum density} $g$ is defined as $%
g=(1/c^{2})S$, i.e., $g$ is $(1/c)$ of the $n-\perp $ part from (\ref{ste})
\begin{equation}
g=-(\varepsilon _{0}/c)\left[ (F\cdot n)\cdot F-(F\cdot n)^{2}n\right] .
\label{ge}
\end{equation}
From $T(n)$ (\ref{ste}) one finds also the expression for \emph{the observer
independent angular-momentum density} $M$
\begin{equation}
M=(1/c)T(n)\wedge x=(1/c)U(n\wedge x)+g\wedge x.  \label{em}
\end{equation}
It has to be emphasized once again that all these definitions are the
definitions of the quantities that are independent of the chosen reference
frame and of the chosen system of coordinates in it; they are all AQs. As I
am aware they are not presented earlier in the literature.

All these quantities can be written in some basis $\left\{ e_{\mu }\right\}
, $ which does not need to be the standard basis, as CBGQs. The field
bivector $F$ can be written as $F=(1/2)F^{\alpha \beta }e_{\alpha }\wedge
e_{\beta }$ where the basis components $F^{\alpha \beta }$ are determined as
$F^{\alpha \beta }=e^{\beta }\cdot (e^{\alpha }\cdot F)=(e^{\beta }\wedge
e^{\alpha })\cdot F$. Then the quantities entering into the expressions for $%
T(n),$ $U, $ $S,$ $g$ and $M$ are $F\cdot F=-(1/2)F^{\alpha \beta }F_{\alpha
\beta },$ $F\cdot n=F^{\alpha \beta }n_{\beta }e_{\alpha },$ $(F\cdot
n)^{2}=F^{\alpha \beta }F_{\alpha \nu }n_{\beta }n^{\nu }$ and $(F\cdot
n)\cdot F=F^{\alpha \beta }F_{\alpha \nu }n_{\beta }e^{\nu }.$ Thence $T(n)$
(\ref{ten1}) becomes
\begin{equation}
T(n)=-(\varepsilon _{0}/2)\left[ (1/2)F^{\alpha \beta }F_{\beta \alpha
}n^{\rho }e_{\rho }+2F^{\alpha \beta }F_{\alpha \rho }n^{\rho }e_{\beta }%
\right] ,  \label{ten2}
\end{equation}
the energy density $U$ (\ref{uen1}) is
\begin{equation}
U=-(\varepsilon _{0}/2)\left[ (1/2)F^{\alpha \beta }F_{\beta \alpha
}+2F^{\alpha \beta }F_{\alpha \rho }n^{\rho }n_{\beta }\right] ,  \label{un2}
\end{equation}
and the Poynting vector $S$ (\ref{po1}) becomes
\begin{equation}
S=-\varepsilon _{0}c\left[ F^{\alpha \beta }F_{\alpha \rho }n^{\rho
}e_{\beta }-F^{\alpha \beta }F_{\alpha \rho }n^{\rho }n_{\beta }n^{\lambda
}e_{\lambda }\right] .  \label{po2}
\end{equation}
In some basis $\left\{ e_{\mu }\right\} $ we can write the stress-energy
vectors $T^{\mu }$ as $T^{\mu }=T(e^{\mu })=(-\varepsilon _{0}/2)Fe^{\mu }F.$
The components of the $T^{\mu }$ represent the energy-momentum tensor $%
T^{\mu \nu }$ in the $\left\{ e_{\mu }\right\} $ basis $T^{\mu \nu }=T^{\mu
}\cdot e^{\nu }=(-\varepsilon _{0}/2)\left\langle Fe^{\mu }Fe^{\nu
}\right\rangle $, which reduces to familiar tensor form
\begin{equation}
T^{\mu \nu }=\varepsilon _{0}\left[ F^{\mu \alpha }g_{\alpha \beta }F^{\beta
\nu }+(1/4)F^{\alpha \beta }F_{\alpha \beta }g^{\mu \nu }\right] .
\label{EMT}
\end{equation}

In the usual Clifford algebra aproach, e.g., [1,2], one makes the space-time
split and considers the energy-momentum density in the $\gamma _{0}$-system
(with the standard basis $\left\{ \gamma _{\mu }\right\} $) $T^{0}=T(\gamma
^{0})=T(\gamma _{0});$ the split $T^{0}\gamma ^{0}=T^{0}\gamma _{0}=T^{00}+%
\mathbf{T}^{0},$ separates $T^{0}$ into an energy density $T^{00}=T^{0}\cdot
\gamma ^{0}$ and a momentum density $\mathbf{T}^{0}=T^{0}\wedge \gamma ^{0}.$
Then from the expression for $T^{\mu }$ and the relations (\ref{hf}) one
finds [1,2] the familiar results for the energy density $T^{00}=(\varepsilon
_{0}/2)(\mathbf{E}_{H}^{2}+c^{2}\mathbf{B}_{H}^{2})$ and the Poyinting
vector $\mathbf{T}^{0}=\varepsilon _{0}(\mathbf{E}_{H}\mathbf{\times }c%
\mathbf{B}_{H}\mathbf{),}$ where the commutator product $A\times B$ is
defined as $A\times B\equiv (1/2)(AB-BA)$. However, as already said, the
space-time split and the introduction of the electric and magnetic fields $%
\mathbf{E}_{H}$ and $\mathbf{B}_{H}$ are not only unnecessary but, as shown
above and in [10,11], they are not equivalent to our general formulation
with AQs. The space-time split is not a Lorentz invariant procedure and if
one wants to use the electric and magnetic fields instead of the bivector
feld $F$ then the decompositions of $F$ into AQs, e.g., $E_{Hv}$\textit{, }$%
B_{Hv}$ from (\ref{bi}), or $E$, $B$ from (\ref{itf}) have to be used and
not the decompositions of $F$ into the observer dependent quantities $%
\mathbf{E}_{H}$, $\mathbf{B}_{H}$ from (\ref{hf}), or $\mathbf{E}_{J}$, $%
\mathbf{B}_{J}$ from (\ref{J1}). \bigskip \medskip

\noindent \textbf{2.7. The\ Local Conservation Laws in the }$F$\textbf{-
Formulation} \textbf{\bigskip }

\noindent It is well-known that from the field equation in the $F$-
formulation (\ref{MEF}) one can derive a set of conserved currents. Thus,
for example, in the $F$- formulation one derives in the standard way that $j$
from (\ref{MEF}) is a conserved current. Simply, the vector derivative $%
\partial $ is applied to the field equation (\ref{MEF}) which yields
\begin{equation*}
(1/\varepsilon _{0}c)\partial \cdot j=\partial \cdot (\partial \cdot F).
\end{equation*}
Using the identity $\partial \cdot (\partial \cdot M(x))\equiv 0$ ($M(x)$ is
a multivector field) one obtains \emph{the local charge conservation law}
\begin{equation}
\partial \cdot j=0.  \label{cjo}
\end{equation}

In the axiomatic formulation [8] the equation (\ref{cjo}), the electric
charge conservation, is axiom 1. We again see that in the axiomatic
formulation with the $F$ field the electric charge conservation is not an
independent axiom but it simply follows from the single axiom for our
theory, the field equation (\ref{MEF}).

In a like manner we find from (\ref{TEF}) ) (which is obtained from (\ref
{MEF})) that
\begin{equation}
\partial \cdot T(n)=0  \label{coti}
\end{equation}
for the free fields. This is a \emph{local energy-momentum conservation law}%
. In the derivation of (\ref{TEF}) we used the fact that $T(a)$ is
symmetric, i.e., that $a\cdot T(b)=T(a)\cdot b.$ Namely using accents the
expression for $T(\partial )$ ($T(\partial )=(-\varepsilon _{0}/2)(F\partial
F)$, where $\partial $ operates to the left and to the right by the chain
rule) can be written as $T(\partial )=\acute{T}(\acute{\partial}%
)=(-\varepsilon _{0}/2)(\acute{F}\acute{\partial}F+F\acute{\partial}\acute{F}%
)=0,$ since in the absence of sources $\partial F=\acute{F}\acute{\partial}%
=0 $ (the accent denotes the multivector on which the derivative acts). Then
from the above mentioned symmetry of $T$ one finds that $\acute{T}(\acute{%
\partial})\cdot a=\partial \cdot T(a)=0$, $\forall \ const.\ a$, which
proves the equation (\ref{coti}).

Inserting the expression (\ref{uis}) for $T(n)$ into the local
energy-momentum conservation law (\ref{coti}) we find
\begin{equation}
(n\cdot \partial )U+(1/c)\partial \cdot S=0.  \label{Poy}
\end{equation}
The relation (\ref{Poy}) is the well-known Poynting's theorem but now
completely written in terms of the observer independent quantities. Let us
introduce the standard basis $\left\{ \gamma _{\mu }\right\} ,$ i.e., an
inertial frame of reference with the Einstein system of coordinates, and in
the $\left\{ \gamma _{\mu }\right\} $ basis we choose that $n=\gamma _{0}$,
or in the component form it is $n^{\mu }(1,0,0,0).$ Then the familiar form
of Poynting's theorem is recovered in such coordinate system
\begin{equation}
\partial U/\partial t+\partial _{i}S^{i}=0,\qquad i=1,2,3.  \label{Poy1}
\end{equation}
It is worth noting that although $U$ (\ref{uen1}) and $S$ (\ref{ste}), taken
separately, are well-defined observer independent quantities, the relations (%
\ref{uis}), (\ref{coti}) and (\ref{Poy}) reveal that only $T(n)$ (\ref{uis}%
), as a whole quantity, i.e., the combination of $U$ and $S,$ enters into a
fundamental physical law, the local energy-momentum conservation law (\ref
{coti}). Thence one can say that only $T(n)$ (\ref{uis}), as a whole
quantity, does have a real physical meaning, or, better to say, a physically
correct interpretation. An interesting example that emphasizes this point is
the case of an uniformly accelerated charge. In the usual (3D) approach to
the electrodynamics ([5]; Jackson, Classical Electrodynamics\textit{,} Sec.
6.8.) the Poynting vector $S$ is interpreted as an energy flux due to the
propagation of fields. In such an interpretation it is not clear how the
fields propagate along the axis of motion since for the field points on the
axis of motion one finds that $S=0$ (there is no energy flow) but at the
same time $U\neq 0$ (there is an energy density). Our approach reveals that
the important quantity is $T(n)$ and not $S$ and $U$ taken separately. $T(n)$
is $\neq 0$ everywhere on the axis of motion and the local energy-momentum
conservation law (\ref{coti}) holds everywhere.

In the same way one can derive the local angular momentum conservation law,
see [1], Space-Time Calculus.\bigskip \bigskip

\noindent \textbf{3. 1-VECTOR\ LAGRANGIAN\ WITH}\ $F$ \bigskip

\noindent In this section we only briefly consider the Lagrangian
formulation. Instead of starting and constructing the whole theory of
electromagnetism using the field equation (\ref{MEF}) one can formulate the
whole theory in terms of the 1-vector Lagrangian written with AQs
\begin{equation}
L=(F\cdot \partial )\cdot (IF)-(IF\cdot \partial )\cdot F-2(IF)\cdot j.
\label{L}
\end{equation}
When $L$ (\ref{L}) is written in terms of CBGQs in the standard basis $%
\{\gamma _{\mu }\}$ it becomes
\begin{eqnarray}
L &=&L_{\alpha }\gamma ^{\alpha },\quad L_{\alpha }=F^{\mu \nu }(\partial
_{\nu }{}^{\ast }F_{\mu \alpha })-  \notag \\
&&^{\ast }F^{\mu \nu }(\partial _{\nu }F_{\mu \alpha })+2{}^{\ast }F_{\alpha
\mu }j^{\mu },  \label{Lu}
\end{eqnarray}
where $L_{\alpha }$ is Sudbery's Lagrangian [21].

The variational principle is applied to
\begin{equation}
S=S_{\alpha }\gamma ^{\alpha }=\left( \int L_{\alpha }\left| d^{4}x\right|
\right) \gamma ^{\alpha };  \label{Sp}
\end{equation}
all four $S_{\alpha }$ should be stationary under variations of $F_{\mu \nu
} $, $j^{\mu }$ being fixed. This leads to Euler-Lagrange equations
\begin{eqnarray}
\lbrack \partial L_{\alpha }/\partial F^{\mu \nu } &-&\partial _{\rho
}(\partial L_{\alpha }/\partial (\partial _{\rho }F^{\mu \nu }))  \notag \\
-(\mu &\longleftrightarrow &\nu )]\gamma ^{\alpha }=0,  \label{EL}
\end{eqnarray}
which, as shown in [21], are equivalent to the full set of the covariant ME (%
\ref{maxco}).

The most important fact in such Lagrangian approach is that \emph{the
interaction term in }(\ref{L}) and (\ref{Lu})\emph{\ is written directly by
means of the measurable electromagnetic field }$F$\emph{, and not, as usual,
in terms of potentials}.\emph{\ }This will have important consequences in
many branches of physics and they will be discussed in future publications.

Using (\ref{L}) and the relations that connect $F$ with different 4D AQs
that represent electric and magnetic fields, e.g., $E$ and $B$ (\ref{itf})
or $E_{Hv}$\textit{\ }$B_{Hv}$ from (\ref{bi}), one can derive the
equivalent Lagrangians with such 4D AQs. \bigskip \bigskip

\noindent \textbf{4. COMPARISON\ WITH\ EXPERIMENTS. THE }

\textbf{EXPLANATION\ OF}\ \textbf{THE TROUTON-NOBLE\ EXPERIMENT\bigskip }

\noindent The usual formulation of special relativity (SR) [12], which deals
with the ``apparent'' transformations, the Lorentz contraction and the
dilatation of time, and the invariant SR from [13] (given in terms of \emph{%
geometric 4D quantities,} abstract tensors) are compared with the
experiments in [14]. It is found in [14] that the usual formulation [12]
shows only an ''apparent'' agreement (not the true one) with the traditional
and modern experiments that test SR, e.g., the Michelson-Morley type
experiments, the ``muon'' experiments, the Ives-Stilwell type experiments,
etc., whereas the invariant SR from [13] is in a \emph{complete agreement}
with all considered experiments. Similarly it is proved in [10] that the
standard transformations of the 3D $\mathbf{E}$ and $\mathbf{B}$ are also
``apparent'' transformations and that they significantly differ from the
correct LT of 4D quantities representing the electric and magnetic fields.
The comparison with experiments on motional electromotive force (emf) given
in the second paper in [10] and also in [11] (the Faraday disk) shows that
the geometric approach with geometric 4D quantities, 4D AQs or equivalently
with 4D CBGQs, and with the LT of the 4D quantities representing the
electric and magnetic fields always agrees with experiments for all
relatively moving observers, whereas it is not the case for the usual
approach with the 3D $\mathbf{E}$ and $\mathbf{B}$ and their standard
transformations. Indeed it is obtained in [11] that the standard formulation
yields different values for the emf of the Faraday disk for relatively
moving inertial observers, see Eqs. (55) and (58) in [11]. (For the
description and the picture of the Faraday disk see, e.g., [22] Chap. 18 or
the first paper in [23].) On the other hand in our geometric approach the
emf is defined as a Lorentz scalar and consequently the same value for that
emf is obtained for all relatively moving inertial frames, see Eqs. (61-63)
in [11].

It is worth noting that in these proofs, e.g., in the second paper in [10],
we could equivalently use the bivector field $F$ through the relation (\ref
{itf}) instead of 1-vectors $E$ and $B$, and similarly in [11].

In this paper we shall discuss the Trouton-Noble experiment [24], see also
[25], comparing the usual explanations with our geometric approach that
explicitly uses AQs, the $F$ field. In the experiment they looked for the
turning motion of a charged parallel plate capacitor suspended at rest in
the frame of the earth in order to measure the earth's motion through the
ether. The explanations, which are given until now (see, e.g., [26-30] and
references therein) for the null result of the experiments [24] ([25]) are
not correct from the invariant SR viewpoint, since they use quantities and
transformations that are not well-defined in the 4D spacetime; e.g., the
Lorentz contraction, the nonelectromagnetic forces of undefined nature, the
standard transformations for the 3D vectors $\mathbf{E}$ and $\mathbf{B}$
and for the torque as the 3D vector, etc.. In all previous treatments it is
correctly found that there is no torque for the stationary capacitor.
However, \emph{the torque is always obtained for the moving capacitor} and
then the above mentioned different explanations are offered for the
existence of another torque which is equal in magnitude but of opposite
direction giving that the total torque is zero. In our approach the
explanation for the null result is very simple and natural; all quantities
are invariant 4D quantities, which means that their values are the same in
the rest frame of the capacitor and in the moving frame. Thus if there is no
torque (but now as a geometric, invariant, 4D quantity) in the rest frame
then the capacitor cannot appear to be rotating in a uniformly moving frame.

Let us discuss the mentioned fundamental difference between the usual
approaches and our geometric approach considering some recent
``explanations'' of the Trouton-Noble paradox. First we examine the
``explanation'' for the null result that is given in [26]. It is shown in
[26] that no turning moment exists both in the rest frame of the capacitor
and in the moving frame. But such correct result is achieved introducing,
together with the electromagnetic forces, some nonelectromagnetic forces,
the forces of constraint, whose physical nature is undefined. Thus, when
using the energy arguments, the null result for the torque in the moving
frame is obtained taking into account not only the electromagnetic energy
but, [26]: ``a contribution of energy from the forces of constraint, due to
work done by these forces during a Lorentz contraction of the system.''
There are several objections to such treatment [26] from the point of view
of the invariant SR and they refer in the same measure to all similar
treatments given in, e.g., [27-29]. These objections are the following:

i) The nonelectromagnetic forces and von Laue's energy current [27]
associated with them are not measurable quantities, see also the discussion
of the Poincar\'{e} stresses and the electromagnetic energy-momentum in
[31]. As Aranoff [32] stated in his severe criticism of von Laue's
explanation: ''The energy current idea of von Laue has to go the way of
phlogiston, and the ether. It is interesting how man has to invent very fine
fluids which carry energy but which are otherwise unobservable.''

ii) The Lorentz contraction is employed in all ``explanations'' [26-29] but,
as shown in [13] and [14], see also [33], the Lorentz contraction has
nothing to do with the LT and moreover it cannot be measured.

iii) The standard transformations of the 3D $\mathbf{E}$ and $\mathbf{B}$
are considered, e.g., [23] and [26-30], to be the LT of these fields.
However it is rigorously proved in [10] that the standard transformations
drastically differ from the correct LT of 4D quantities representing the
electric and magnetic fields.

iv) The transformations of the components of a 3D force and torque are
commonly used in the mentioned ``explanations,'' but the correct LT always
refer to the 4D quantities. In the invariant SR the theoretical and \emph{%
experimental} meaning is attributed only to geometric 4D quantities and not
to their parts. Different relatively moving inertial 4D observers can
compare only 4D quantities since they are connected by the LT.

In the recent paper [30] it is argued that the Trouton-Noble paradox is
resolved once the electromagnetic momentum of the moving capacitor is
properly taken into account. First it is obtained that there is a mechanical
3D torque on the moving capacitor and then it is shown that the rate of
change of the angular electromagnetic field momentum associated with the
moving capacitor completely balances that mechanical torque. We want to show
that the appearance of the 3D torque on the moving capacitor in [30] is a
consequence of the above mentioned objection iv) and another one:

v) The use of the principle of relativity for physical laws that are
expressed by 3D quantities.

It will be seen below that in the geometric approach with 4D quantities the
torque will not appear for the moving capacitor if it does not exist for the
stationary capacitor. Thus, actually, the consideration with 4D quantities
and their LT will reveal that there is no need at all either for the
nonelectromagnetic forces and their torque, [26-29], or for the angular
electromagnetic field momentum and its rate of change, i.e., its torque,
[30]. Therefore we shall examine in more detail the calculation of the
torque that is presented in [30], but not of the angular electromagnetic
field momentum. In the rest frame of a thin parallel-plate capacitor, the $%
S^{\prime }$ frame, there is no torque. Then it is assumed in [30] that in
the $S$ frame the capacitor moves with uniform velocity $\mathbf{V}$ (the 3D
vector) in the positive direction of the $x^{1}$ - axis. (Fig. 1. from [30]
is actually a projection onto the hypersurface $t^{\prime }=const.$, which
means that $x$, $y$ and $\Theta $ from that Fig.1. would need to be denoted
as $x^{\prime 1}$, $x^{\prime 2}$ and $\Theta ^{\prime }$ respectively.) In
the $S^{\prime }$ frame $A$ denotes the surface area of the capacitor's
plates, $a$ is the distance between the capacitor's plates and $\Theta
^{\prime }$ is the angle between the line joining the axis of rotation
(i.e., the middle of the negative plate) with the middle of the positive
plate and the $x^{\prime 2}$ axis. That line is taken to be in the $%
x^{\prime 1}$, $x^{\prime 2}$ plane (see Fig. 1. in [30]). The torque (the
3D vector) experienced by the moving capacitor is determined by using
``relativistic'' (my quotation-marks) transformation equations for the
torque. These ``relativistic'' transformation equations for the 3D torque
given in [30] are
\begin{equation}
N_{1}=N_{1}^{\prime }/\gamma ,\ N_{2}=N_{2}^{\prime
}+(V^{2}/c^{2})r_{1}^{\prime }K_{cl.3}^{\prime },\ N_{3}=N_{3}^{\prime
}-(V^{2}/c^{2})r_{1}^{\prime }K_{cl.2}^{\prime },  \label{s}
\end{equation}
where $\gamma =(1-V^{2}/c^{2})^{-1/2}$, $K_{cl.i}^{\prime }$ are the
components of the 3D force acting on the positive plate of the stationary
capacitor, and $r_{i}^{\prime }$ are the components of the lever arm joining
the axis of rotation with the point of application of the resultant force,
i.e., the midpoint of the positive plate, see Fig. 1. in [30]. (The
equations (\ref{s}) are the equations (1)-(3) in [30].) As already said the
3D torque on the stationary capacitor is zero, $N_{i}^{\prime }=\varepsilon
_{ijk}r_{j}^{\prime }K_{cl.k}^{\prime }=0$. (Note that in this equation for $%
N_{i}^{\prime }$ and in (\ref{s}) we have used the same notation as in (\ref
{sko1}), i.e., the components of the 3D quantities are written with lowered
(generic) subscripts, since they are not the spatial components of the 4D
quantities. This refers to the third-rank antisymmetric $\varepsilon $
tensor too.) Taking into account that $N_{i}^{\prime }=0$ and $%
K_{cl.3}^{\prime }=0$ Jefimenko [30] finds that $N_{3}$ component is
different from zero
\begin{equation}
N_{3}=-(V^{2}/c^{2})r_{1}^{\prime }K_{cl.2}^{\prime }.  \label{k5}
\end{equation}
This result is commented in [30] in the following way: ``We have thus
obtained a paradoxical result: contrary to the relativity principle,
although our stationary capacitor experiences no torque, the same capacitor
moving with uniform velocity along a straight line appears to experience a
torque. What makes this result especially surprising is that we have arrived
at it by using relativistic transformations that are based on the very same
relativity principle with which they now appear to conflict.''

Let us examine the calculation leading to (\ref{s}) and (\ref{k5}) and the
above quoted statements. First $N_{i}^{\prime }$ is defined by means of the
3D quantities. Then the transformation equations (\ref{s}) for the
components of the 3D torque are derived considering that the transformations
of the components of the 3D force are the LT. Since in $S^{\prime }$ the
capacitor is at rest the mentioned transformations of the components of the
3D force are
\begin{equation}
K_{cl.1}=K_{cl.1}^{\prime },\ K_{cl.2}=K_{cl.2}^{\prime }/\gamma ,\
K_{cl.3}=K_{cl.3}^{\prime }/\gamma .  \label{efi}
\end{equation}
It is assumed in [30], as in many other papers including [26-29], that the
transformations (\ref{efi}) (and similarly for (\ref{s})) are the
relativistic transformations, i.e., the LT, that are based on the principle
of relativity. Such opinion implicitly supposes that 3D quantities, their
transformations and physical laws written in terms of them are physically
real in the 4D spacetime and in agreement with the principle of relativity.
Actually such opinion prevails already from Einstein's fundamental work on
SR [12].

The approach of the invariant SR [13-15] and [10,11] is completely
different. There, as already explained, the physical reality in the 4D
spacetime is attributed only to geometric 4D quantities, AQs or CBGQs, their
LT and physical laws written in terms of them. The principle of relativity
is automatically included in such formulation. Thence in the 4D spacetime we
are dealing with the Lorentz force $K=(q/c)F\cdot u$, where $u$ is the
velocity 1-vector of a charge $q$. The torque, as a 4D AQ, is defined as the
bivector
\begin{equation}
N=r\wedge K,\ r=x_{P}-x_{O},  \label{or}
\end{equation}
where $r$ is 1-vector associated with the lever arm, $x_{P}$ and $x_{O}$ are
the position 1-vectors associated with the spatial point of the axis of
rotation and the spatial point of application of the force $K$, $P$ and $O$
are the events whose position 1-vectors are $x_{P}$ and $x_{O}$.

In general the proper velocity $u$ for a point particle is $u=dx/d\tau $, $%
\tau $ is the proper time, $p$ is the proper momentum $p=mu$, the proper
angular momentum of a particle is the bivector $L=x\wedge p$ and the torque $%
N$ about the origin is the bivector $N=dL/d\tau =x\wedge K$, where in this
relation $K$ is an arbitrary force 1-vector. When $K$ is written as a CBGQ
in the standard basis $\{\gamma _{\mu }\}$ then its components are $K^{\mu
}=(\gamma _{u}K_{cl.i}V_{i}/c,\gamma _{u}K_{cl.1},\gamma _{u}K_{cl.2},\gamma
_{u}K_{cl.3})$, and the components of $u$ in the $\{\gamma _{\mu }\}$ basis
are $u^{\mu }=(\gamma _{u}c,\gamma _{u}V_{1},\gamma _{u}V_{2},\gamma
_{u}V_{3})$. $\gamma _{u}=(1-V^{2}/c^{2})^{-1/2}$, $K_{cl.i}$ are components
of the 3D force and $V_{i}$ are components of the 3D velocity. We see that
only when the considered particle is at rest, i.e., $V_{i}=0$, $\gamma
_{u}=1 $ and consequently $u^{\mu }=(c,0,0,0)$, then $K^{\mu }$ contains
only the components $K_{cl.i}$, i.e., $K^{\mu
}=(0,K_{cl.1},K_{cl.2},K_{cl.3})$. However even in that case $u^{\mu }$ and $%
K^{\mu }$ are the components of geometric \emph{4D} \emph{quantities }$u$
and $K$ in the $\{\gamma _{\mu }\}$ basis and not the components of some
\emph{3D quantities} $\mathbf{V}$ and $\mathbf{K}_{cl.}$. The LT correctly
transform the whole 4D quantity, which means that there is no physical sense
in such transformations like (\ref{efi}) and (\ref{s}); these
transformations are not relativistic and they are not based on the principle
of relativity. All conclusions derived from such relations as are (\ref{efi}%
) and (\ref{s}) have nothing in common with SR as the theory of the 4D
spacetime.

When $N$ (\ref{or}) is written as a CBGQ in the standard basis $\{\gamma
_{\mu }\}$ then it becomes $N=(1/2)N^{\mu \nu }\gamma _{\mu }\wedge \gamma
_{\nu }$, where $N^{\alpha \beta }=\gamma ^{\beta }\cdot (\gamma ^{\alpha
}\cdot N)=r^{\alpha }K^{\beta }-r^{\beta }K^{\alpha }$. Let us prove that $N$
is zero, $N=0$, in the rest frame of the capacitor, here the $S^{\prime }$
frame. In that frame we choose that $r^{\prime 0}=x_{P}^{\prime
0}-x_{O}^{\prime 0}=0$. The system of coordinates is chosen in such a way
that $r^{\prime 3}=0$ (as in Fig. 1. [30]) giving that $r^{\prime \mu
}=(0,r_{1}^{\prime },r_{2}^{\prime },0)$. Further, since in $S^{\prime }$ we
have stationary capacitor, $V_{i}=0$, $\gamma _{u}=1$, and from the chosen
system of coordinates we conclude that $K_{cl.3}^{\prime }=0$, which yields $%
K^{\prime \mu }=(0,K_{cl.1}^{\prime },K_{cl.2}^{\prime },0)$. Thence we find
that $N^{\prime i0}=N^{\prime 13}=N^{\prime 23}=0$ and only remains $%
N^{\prime 12}=r^{\prime 1}K^{\prime 2}-r^{\prime 2}K^{\prime 1}$. Let us
prove that $N^{\prime 12}$ is also zero. We shall use the result (\ref{fp})
obtained in Sec. 2.4 for $F$ of a flat sheet with the constant surface
charge density $\sigma $, then also the chosen system of coordinates, i.e.,
Fig. 1. from [30], and the relation (\ref{itf}). Taking that in the relation
(\ref{itf}) the velocity $v$ of the observers in the $S^{\prime }$ frame is $%
v=c\gamma _{0}^{\prime }$, i.e., that the $S^{\prime }$ frame is the frame
of ``fiducial'' observers, we have that $F^{\prime i0}=E^{\prime i}$ and all
other $F^{\prime \mu \nu }$ are zero. Then we can employ the discussion from
Sec. 2. in [30]. In that discussion the electric field (as a 3D vector) and
the 3D force $\mathbf{K}_{cl.}^{\prime }$ are determined. The electric field
is produced by the negative plate of the capacitor at the location of the
positive plate. The 3D force $\mathbf{K}_{cl.}^{\prime }$ acting on the
positive plate is along the line joining the axis of rotation (i.e., the
middle of the negative plate) with the middle of the positive plate. All
this together yields that $F^{\prime 10}=(\sigma /2\varepsilon _{0})\sin
\Theta ^{\prime }$, $F^{\prime 20}=-(\sigma /2\varepsilon _{0})\cos \Theta
^{\prime }$ and $K^{\prime }=(\sigma A)(F^{\prime 10}\gamma _{1}^{\prime
}+F^{\prime 20}\gamma _{2}^{\prime })$ and also $r^{\prime 1}=-a\sin \Theta
^{\prime }$, $r^{\prime 2}=a\cos \Theta ^{\prime }$, where, as already said,
$A$ is the surface area of the capacitor's plates and $a$ is the distance
between the capacitor's plates. Thence we find that $N^{\prime 12}=(\sigma
^{2}A/2\varepsilon _{0})a(\sin \Theta ^{\prime }\cos \Theta ^{\prime }-\sin
\Theta ^{\prime }\cos \Theta ^{\prime })=0$. Thus all $N^{\prime \alpha
\beta }$ are zero in the $S^{\prime }$ frame in which the capacitor is at
rest. Since the CBGQ $(1/2)N^{\prime \mu \nu }\gamma _{\mu }^{\prime }\wedge
\gamma _{\nu }^{\prime }$ is an invariant quantity upon the passive LT we
have proved that not only the components $N^{\prime \alpha \beta }$ are zero
but at the same time that the whole torque $N$ is zero
\begin{equation}
N=(1/2)N^{\prime \mu \nu }\gamma _{\mu }^{\prime }\wedge \gamma _{\nu
}^{\prime }=(1/2)N^{\mu \nu }\gamma _{\mu }\wedge \gamma _{\nu }=0.
\label{en1}
\end{equation}
Thence the torque is zero not only for the stationary capacitor but for the
moving capacitor as well. We see that in the approach with the geometric 4D
quantities there is no Trouton-Noble paradox. \bigskip \bigskip

\noindent \textbf{5. DISCUSSION\ AND\ CONCLUSIONS\bigskip }

\noindent The aim of this work is to present an axiomatic, geometric
approach to electromagnetism in which the primary quantity is the
electromagnetic field $F$ as an observer independent 4D quantity. The whole
theory is deduced from only one axiom: the field equation for $F$ (\ref{MEF}%
). This formulation with the $F$ field is a self-contained, complete and
consistent formulation that does not make use either electric and magnetic
fields or the electromagnetic potential $A$. Such approach conceptually
differs from all previous approaches in several respects.

First, it places the electromagnetic field $F$ in the centre of the
theoretical formulation and not, as usual, the 3D $\mathbf{E}$ and $\mathbf{B%
}$.

Second, the bivector field $F$ is considered to have an independent physical
reality as a measurable 4D quantity, see particularly the end of Sec. 2.5
and Sec. 4.

Third, the whole theory is manifestly Lorentz \emph{invariant}; it deals
only with 4D AQs or 4D CBGQs and the space-time split, i.e., the foliation
of the spacetime, is not introduced anywhere.

Fourth, the connection with the usual picture that deals with electric and
magnetic fields is given by the relations (\ref{itf}) or (\ref{he}). All
quantities in these relations are 4D AQs in contrast to the common
decompositions of $F$, e.g., from [1-3], into observer dependent quantities $%
\mathbf{E}_{H},$ $\mathbf{B}_{H}$ [1,2], or $\mathbf{E}_{J},$ $\mathbf{B}%
_{J} $ [3]. Every relation with $F$ can be transformed to to the
corresponding relation with electric and magnetic fields as 4D AQs using the
mentioned equations (\ref{itf}) or (\ref{he}); see, for example, Sec. 2.3 in
which the electromagnetic field of a point charge is considered.

Fifth, many new results are obtained here, which are not yet presented in
the literature. However even in the cases when we used the results already
presented in the literature, particularly in [1-3], these results are
interpreted and explained in such a way to be in agreement with our
axiomatic formulation given in terms of 4D AQs or 4D CBGQs and without any
use of the space-time split. This differs from all previous approaches,
e.g., [1-3] and [7,8].

\emph{The observer independent expressions }for the stress-energy vector $%
T(n),$ the energy density $U,$ the Poynting vector $S,$ the momentum density
$g,$ the angular-momentum density $M$ and the Lorentz force $K$ are derived
from the field equation (\ref{MEF}) and presented in Sec. 2.6 in this paper.
Then the second quantization procedure, and the whole quantum
electrodynamics, can be constructed using these geometric, invariant,
quantities $F$, $T(n)$, $U$, $S$, $g$ and $M.$ Note that the standard
covariant approaches to quantum electrodynamics, e.g., [34], usually deal
with the component form (in the specific, i.e., the Einstein system of
coordinates) of the electromagnetic 4-potential $A$ (thus requiring the
gauge conditions too) and not with geometric quantities, AQs or CBGQs. The
local conservation laws are also directly derived from the field equation (%
\ref{MEF}) and written in an invariant way in Sec. 2.7. The observer
independent integral field equation (\ref{dk}) corresponding to the field
equation (\ref{MEF}) is quoted and discussed in Sec. 2.4. In Sec. 3 we have
constructed 1-vector Lagrangian $L$ (\ref{L}), corresponding to the field
equation (\ref{MEF}), with a specific feature that the interaction term is
written in terms of $F$ and not, as usual, in terms of potential $A$. When
that $L$ (\ref{L}) is written in the standard basis $\{\gamma _{\mu }\}$ it
becomes Sudbery's Lagrangian [21]. Such form of the Lagrangian suggests that
in the classical electromagnetism, contrary to the generally accepted
opinion, the interaction term can be expressed exclusively by means of
measurable quantities, either $F$, or electric and magnetic fields as 4D
geometric quantities when, e.g., the relations (\ref{itf}) or (\ref{he}) are
used. The consequences to the quantum mechanics will be examined elsewhere.

Particularly it has to be emphasized that the observer independent approach
to the relativistic electrodynamics that is presented in this paper is in a
complete agreement with existing experiments that test special relativity,
which is not the case with the usual approaches. This is shown in detail in
Sec. 4 for the Trouton-Noble experiment.

Furthermore we note that all observer independent quantities introduced here
and the field equations written in terms of them hold in the same form both
in the flat and curved spacetimes. The formalism presented here will be the
basis for the formulation of quantum electrodynamics and, more generally, of
the quantum field theory that exclusively deals with AQs or CBGQs.\bigskip
\bigskip

\noindent \textbf{ACKNOWLEDGMENTS\bigskip }

\noindent I thank Tony Sudbery for sending me his calculation. Thanks also
Larry Horwitz for his continuous support and useful comments, Alex Gersten,
Bill Schieve, Matej Pav\v{s}i\v{c} (and other participants of the IARD 2004
Conference) for interesting discussions, and Zbigniew Oziewicz, Valeri
Dvoeglazov and Bernard Jancewicz for useful correspondence. \bigskip \bigskip

\noindent \textbf{REFERENCES}\bigskip

\noindent 1. D. Hestenes, \textit{Space-Time Algebra }(Gordon and Breach,
New York, 1966);

\textit{Space-Time Calculus; }available at: http://modelingnts.la.
asu.edu/evolution.

html; \textit{New Foundations for Classical Mechanics }(Kluwer, Dordrecht,

1999) 2nd. edn.; \textit{Am. J Phys.} \textbf{71}, 691 (2003).

\noindent 2. C. Doran, and A. Lasenby, \textit{Geometric algebra for
physicists }

(Cambridge University, Cambridge, 2003).

\noindent 3. B. Jancewicz, \textit{Multivectors and Clifford Algebra in
Electrodynamics}

(World Scientific, Singapore, 1989).

\noindent 4. D. Hestenes and G. Sobczyk, \textit{Clifford Algebra to
Geometric Calculus}

(Reidel, Dordrecht, 1984).

\noindent 5. J.D. Jackson, \textit{Classical Electrodynamics} (Wiley, New
York, 1977) 2nd

edn.. L.D. Landau and E.M. Lifshitz, \textit{The Classical Theory of Fields }

(Pergamon, Oxford, 1979) 4th edn.. C.W. Misner, K.S.Thorne, and J.A.

Wheeler, \textit{Gravitation} (Freeman, San Francisco, 1970). W.G.T.V.
Rosser,

\textit{Classical Electromagnetism via Relativity} (Plenum, New York, 1968).

\noindent 6. A. Einstein, \textit{Ann. Physik} \textbf{49,} 769 (1916), tr.
by W. Perrett and G.B.

Jeffery, in \textit{The Principle of Relativity }(Dover, New York, 1952).

\noindent 7. Yu.N. Obukhov and F.W. Hehl, \textit{Phys. Lett. A} \textbf{311}%
, 277 (2003).

\noindent 8. F.W. Hehl, Yu.N. Obukhov, \textit{Foundations of Classical
Electrodynamics}

(Birkh\"{a}user, Boston, MA, 2003). F.W. Hehl, Yu.N. Obukhov and

G.F. Rubilar, physics/9907046. F.W. Hehl, Yu.N. Obukhov, physics/0005084.

\noindent 9. J.J. Cruz Guzm\'{a}n, Z. Oziewicz, \textit{Bull. Soc. Sci.
Lett. L\'{o}d\'{z}} \textbf{53}, 107 (2003).

\noindent 10. T. Ivezi\'{c}, Found. Phys. \textbf{33}, 1339 (2003); physics/%
\textit{0411166 (}to be

published in\textit{\ }Foundations of Physics Letters\textit{)}.

\noindent 11. T. Ivezi\'{c}, physics/\textit{0409118 v2 (}to be published in%
\textit{\ }Foundations of Physics).

\noindent 12. A. Einstein, \textit{Ann. Physik.} \textbf{17}, 891 (1905),
tr. by W. Perrett and G.B.

Jeffery, in \textit{The Principle of Relativity} (Dover, New York, 1952).

\noindent 13. T. Ivezi\'{c}, \textit{Found. Phys.} \textbf{31}, 1139 (2001).

\noindent 14. T. Ivezi\'{c}, \textit{Found. Phys. Lett.} \textbf{15}, 27
(2002); physics/0103026; physics/

0101091.

\noindent 15. T. Ivezi\'{c}, hep-th/0207250; hep-ph/0205277.

\noindent 16. M. Riesz, \textit{Clifford Numbers and Spinors}, Lecture
Series No. 38

(The Institute for Fluid Dynamics and Applied Mathematics,

University of Maryland, 1958).

\noindent 17. D. Hestenes, in \textit{Clifford Algebras and their
Applications in }

\textit{Mathematical Physics}, F. Brackx et al, Eds. (Kluwer, Dordrecht,
1993).

\noindent 18. T. Ivezi\'{c} and Lj. \v{S}kovrlj, unpublished results. Lj. \v{%
S}kovrlj, \textit{Thesis} (2002)

(in Croatian).

\noindent 19. R.M. Wald, \textit{General Relativity} (Chicago University,
Chicago,

1984). M. Ludvigsen, \textit{General Relativity,} \textit{A Geometric
Approach }

(Cambridge University, Cambridge, 1999). S. Sonego and

M.A. Abramowicz, \textit{J. Math. Phys.} \textbf{39}, 3158 (1998). D.A. T.
Vanzella,

G.E.A. Matsas, H.W. Crater, \textit{Am. J. Phys.} \textbf{64}, 1075 (1996).

\noindent 20. A.T. Hyman, \textit{Am. J. Phys.} \textbf{65}, 195 (1997). G.
M\~{u}noz, \textit{Am. J. Phys.}

\textbf{65}, 429 (1997).

\noindent 21. A. Sudbery, \textit{J. Phys. A: Math. Gen.} \textbf{19,}
L33-36 (1986).

\noindent 22. W.K.H. Panofsky and M. Phillips, \textit{Classical electricity
and magnetism}

(Addison-Wesley, Reading, Mass., 1962) 2nd edn.

\noindent 23. L. Nieves, M. Rodriguez, G. Spavieri and E. Tonni, \textit{%
Nuovo Cimento B}

\textbf{116}, 585 (2001). G. Spavieri and G.T. Gillies, \textit{Nuovo
Cimento B}

\textbf{118}, 205 (2003).

\noindent 24. F.T. Trouton and H.R. Noble, \textit{Philos. Trans. R. Soc.
London Ser. A}

\textbf{202}, 165 (1903).

\noindent 25. H.C. Hayden, \textit{Rev. Sci. Instrum.} \textbf{65}, 788
(1994).

\noindent 26. A.K. Singal, \textit{Am. J. Phys.} \textbf{61}, 428 (1993).

\noindent 27. M. von Laue, \textit{Phys. Zeits.} \textbf{12}, 1008 (1911).

\noindent 28. W. Pauli, \textit{Theory of Relativity} (Pergamon, New York,
1958)

\noindent 29. S. A. Teukolsky, \textit{Am. J. Phys.} \textbf{64}, 1104
(1996).

\noindent 30. O.D. Jefimenko, \textit{J. Phys. A: Math. Gen.} \textbf{32,}
3755 (1999).

\noindent 31. T. Ivezi\'{c}, \textit{Found. Phys. Lett.} \textbf{12}, 105
(1999)

\noindent 32. S. Aranoff, \textit{Nuovo Cimento B} \textbf{10}, 155 (1972).

\noindent 33. T. Ivezi\'{c}, \textit{Found. Phys. Lett.} \textbf{12}, 507
(1999).

\noindent 34. J.D. Bjorken and S.D. Drell, \textit{Relativistic Quantum Field%
} (McGraw-Hill,

New York, 1964). F. Mandl and G. Shaw, \textit{Quantum Field Theory} (Wiley,

New York, 1995). S. Weinberg, \textit{The} \textit{Quantum Theory of}

\textit{Fields, Vol. I Foundations }(Cambridge University, Cambridge,

1995).

\end{document}